\documentstyle[12pt]{article}
\newcommand{\real}{{\rm I\kern-.17em R}}

\newcommand{\au}{
\begin{picture}(11,3)(-2,-2)
\put(2,8){$\scriptscriptstyle{1}$}
\put(0,-2){A}
\end{picture}}

\newcommand{\ad}{
\begin{picture}(11,3)(-2,-2)
\put(2,8){$\scriptscriptstyle{2}$}
\put(0,-2){A}
\end{picture}}

\newcommand{\bu}{
\begin{picture}(11,3)(-2,-2)
\put(2,8){$\scriptscriptstyle{1}$}
\put(0,-2){B}
\end{picture}}

\newcommand{\bd}{
\begin{picture}(11,3)(-2,-2)
\put(2,8){$\scriptscriptstyle{2}$}
\put(0,-2){B}
\end{picture}}

\newcommand{\cu}{
\begin{picture}(11,3)(-2,-2)
\put(2,8){$\scriptscriptstyle{1}$}
\put(0,-2){C}
\end{picture}}

\newcommand{\cd}{
\begin{picture}(11,3)(-2,-2)
\put(2,8){$\scriptscriptstyle{2}$}
\put(0,-2){C}
\end{picture}}

\newcommand{\fu}{
\begin{picture}(11,3)(-2,-2)
\put(0,6){$\scriptscriptstyle{1}$}
\put(-2,-2){$\varphi$}
\end{picture}}

\newcommand{\fd}{
\begin{picture}(11,3)(-2,-2)
\put(0,6){$\scriptscriptstyle{2}$}
\put(-2,-2){$\varphi$}
\end{picture}}

\newcommand{\Au}{
\begin{picture}(11,3)(-2,-2)
\put(2,6){$\scriptscriptstyle{1}$}
\put(0,-2){a}
\end{picture}}

\newcommand{\Ad}{
\begin{picture}(11,3)(-2,-2)
\put(2,6){$\scriptscriptstyle{2}$}
\put(0,-2){a}
\end{picture}}

\newcommand{\Bu}{
\begin{picture}(11,3)(-2,-2)
\put(2,8){$\scriptscriptstyle{1}$}
\put(0,-2){b}
\end{picture}}

\newcommand{\Bd}{
\begin{picture}(11,3)(-2,-2)
\put(2,8){$\scriptscriptstyle{2}$}
\put(0,-2){b}
\end{picture}}

\newcommand{\Fu}{
\begin{picture}(11,3)(-2,-2)
\put(1,7){$\scriptscriptstyle{1}$}
\put(-2,-2){$\phi$}
\end{picture}}

\newcommand{\Fd}{
\begin{picture}(11,3)(-2,-2)
\put(1,7){$\scriptscriptstyle{2}$}
\put(-2,-2){$\phi$}
\end{picture}}

\newcommand{\p}{
\begin{picture}(11,3)(-2,-2)
\put(2,8){$\scriptscriptstyle{1}$}
\put(0,-2){P}
\end{picture}}

\begin{document}

\baselineskip=22pt plus 0.2pt minus 0.2pt
\lineskip=22pt plus 0.2pt minus 0.2pt
\begin{center}
\vspace*{1cm}
\LARGE
In Search of Local Degrees of Freedom\\
in Quadratic Diff-invariant Lagrangians.\\

\vspace*{1.5cm}

\large                                                                        

J.\ Fernando\ Barbero\ G.,\\and\\
Eduardo J. S. Villase\~nor

\vspace*{1.5cm}

\normalsize
{\it Escuela Superior de Ingenier\'{\i}a Industrial,\\
Universidad Europea \\
Urb. El Bosque, C/ Tajo s/n\\
Villaviciosa de Od\'on, Madrid, 28670\\
Spain\\\hspace{5mm}\\}

\vspace{.3in}
September 20, 1999\\
\vspace{.3in}
ABSTRACT
\end{center}

We show that local diff-invariant free field theories in four 
spacetime dimensions do not have local degrees of freedom.

\vspace*{1cm}
\noindent PACS number(s): 04.20.Cv, 04.20.Fy

\pagebreak

\setcounter{page}{1}        

\section{Introduction}

There is a clear consensus nowadays about the most promising 
strategies to find a consistent theory of quantum gravity. 
An important majority of the leading theoretical physicists
(and their followers) consider that superstring  theories and their derivatives,
especially M-theory\footnote{ See \cite{Duff} for a down to earth presentation on the subject or \cite{Schwarz}, \cite{Pol} and references therein for a more complete treatment.}
hold  the key to understanding general relativity
in the quantum regime. This does not mean that 
there are no other potentially successful approaches to tackle this problem such as perturbative quantum gravity, QFT in curved backgrounds, Euclidean quantum gravity, Regge calculus and lattice techniques, twistor theory, non-commutative geometry, non-perturbative quantum gravity and many others \cite{Gibbs}. Some of the most recent approaches, such as Ashtekar's non-perturbative formulation of gravity \cite{ASH}
and the general 
setting provided by the loop variables formalism to deal with diff-invariant
theories \cite{LOOP}
have received a lot of attention in recent years as very promising ways to tame the deep problems presented by the quantization of general relativity. 
There is also a clear consensus about the 
failure of perturbative formulations for gravity although there 
have been some recent and interesting papers on the subject \cite{Be}; it is thus curious
to look at a series of papers \cite{EW1}, \cite{DMY} in the late eighties and early
nineties\footnote{One of us (J.F.B.G.) is grateful to
Abhay Ashtekar for drawing his attention to these papers.} that give a novel understanding 
on this issue. In \cite{EW1} Witten shows that  2 + 1 gravity
has a renormalizable perturbation expansion 
and gives the following
explanation about the non-renormalizability in the 3+1 
dimensional case:

\bigskip

``It is amusing to think about $3+1$ dimensional gravity
from this point of view. The lagrangian is of the general form
$$
I_{(4)}\sim \int e\wedge e\wedge (d \omega+\omega\wedge \omega)\quad. \quad\quad
(3.13)
$$
If one hopes for ``power-counting renormalizability'', one
needs to assign dimension one to both $e$ and $\omega$, so
that every term in eq. (3.13) is of dimension four. (Again, this
is in contrast to the fact that the metric and vierbein are 
usually considered to have dimension zero). As $e$ and $\omega$ 
have positive dimension, the short-distance limit must
have $e=\omega=0$. The problem is now that as eq. (3.13) has no
quadratic term in an expansion around $e=\omega=0$, one cannot make sense of the
``unbroken phase'' that should govern the short-distance
behavior; that is the essence of the unrenormalizability of quantum
gravity in four dimensions.''

\bigskip

	Following this suggestion Deser, McCarthy and Yang  \cite{DMY}
analyzed
the same problem by using a Palatini action. The advantage
of their approach  is that while preserving power counting renormalizability
in $D=1,\dots,4$ the action has a quadratic term in an expansion about
zero field value. The main result
of that paper, in the 3+1 dimensional context, was to show that 
non-renormalizability
could be traced back to the mismatch
between the symmetries of the full action and those of the kinetic term.
Specifically, they showed that the quadratic part of the action
had more independent gauge symmetries that the full
action and hence one could not render the kinetic part invertible by
using only the maximum number of gauge fixing conditions allowed
by the symmetry of the full
Lagrangian. The authors of that paper wonder

\bigskip

``Whether a viable local modification of gravity
which exploits this ``near miss'' exists is an open question''.

\bigskip

	These are several conceivable ways to advance in this
direction such as looking for other actions for general relativity
--with quadratic terms that can hopefully be inverted after gauge fixing--
or modify the theory to provide it with kinetic terms with the desired 
properties.

A nice way to do this would be to add a diff-invariant quadratic kinetic
term written in terms of the fields appearing in the action
\begin{equation}
S=\int e\wedge e\wedge F(\omega)\quad,
\label{001}
\end{equation}
where $e$ is a tetrad one--form and $F(w)$ the curvature of a $SO(3,1)$
spin connection\footnote{Notice that this is something
conceptually similar to the introduction of higher
derivative terms but in the opposite direction.}
$\omega$. If this is to succeed the
quadratic  part of the action should have, at least, two physical 
degrees of freedom because if it had less then the kinetic term 
would have more gauge symmetry than the full action and hence by gauge 
fixing the symmetries present in  (\ref{001}) one would not  
get  an invertible kinetic term.
Also some matching of the symmetries of 
$S$ and the kinetic term should be imposed. With this philosophy in mind
one is naturally led to pose the following question:

{\it 
Can a diff-invariant quadratic Lagrangian have local degrees of freedom 
in four dimensions?}

The purpose of this paper is to answer this question. We want
to stress that we are taking the 
point of view of Witten in \cite{EW1} that with fields
of dimension $+1$, the perturbative  expansions
should be performed about a zero field background. We will
not consider expansions about non-trivial backgrounds.

\bigskip

The paper is organized as follows. After this introduction we devote section 
II to the construction of the most general local 
quadratic diff-invariant action in four dimensions (under
some mild restrictions). The field equations derived from this 
action are linear and can be completely solved, this is 
the purpose of section III where we do this in a systematic
way. The solutions depend on a series of arbitrary elements of two 
different types: some
of them correspond to the gauge symmetries present 
in the action whereas some others label non gauge equivalent  solutions.
 In order
to disentangle their role one must use the symplectic structure,
this is done in section IV. It is very important to realize 
that without the  use of the symplectic 2--form
it is not possible to tell which of the arbitrary 
parameters are gauge and which are not. Once the symplectic
structure is obtained the identification of the physical 
degrees of freedom and gauge symmetries
is straightforward. We discuss this  and give some
examples in section V. We end the paper with several
comments and our conclusions (section VI).
 Some details of the computations are left to the appendices.

\section{ The Action}

	As we have already stated in the introduction, the main goal 
of this paper is to search for local degrees of freedom
in quadratic diff-invariant actions, so our first task will
be to write the most general action of this type
under some restrictions. In particular 
we will demand
\begin{itemize}
\item[i.] Absence of background structures: All the fields appearing 
in the action must be treated as dynamical. 

\item[ ii.] Locality: The action must be local in the fields used to
define it. This is arguably the most stringent condition
that we impose.

\item[ iii.] The action must be, at most, quadratic in all the fields.

\end{itemize}   

	The previous assumptions strongly constrain the 
possible form of the action. By combining the absence 
of background structures with its quadratic character we arrive at
the conclusion that the only derivative operator that can
possible appear is the exterior derivative acting on differential forms.
Covariant derivatives of other types of tensor fields  cannot 
be used as they would involve quadratic terms that would 
force them to appear as total divergences in the action\footnote{We
discard them by working with manifolds without boundary, we discuss the introduction of boundaries in appendix 1.}. For the same
reason derivatives can only
act on differential forms (no other types of tensors can appear).  
Purely quadratic terms involving no derivatives
can also be introduced; in particular, pairs of tensor fields
with matching covariant and contravariant indices and total density weight 
$+1$. Let us consider ${\cal M}$,  a four dimensional, orientable differentiable manifold
with the topology of $\real \times \Sigma$
where $\Sigma$ is a three-dimensional compact, orientable manifold without boundary. 
We make a distinction between two types of fields; type 1 fields are those   on which the derivative operator acts
(either directly or after integration by parts); type 2 fields
are not acted upon by the exterior differential operator
but couple to type 1 fields. 
The dynamical (real) fields  that will appear in the first part of
the action that we introduce below are $\fu$, $\fd$, $\au$, $\ad$, $\bu$, $\bd$, 
$\cu$, $\cd$ and  $D$. The $\varphi$ fields are 0-forms, 
$A$ fields are 1-forms, $B$ fields are 2-forms, 
$C$ fields are 3-forms and $D$ is a 4-form; they may carry internal indices, so we use the convention 
of taking them as column vectors and use the transpose
(that we denote as $\dagger$ even though we are dealing with
real fields) whenever necessary. $\Omega_{11}\,$, $\Omega_{12}\,$,
$\Omega_{22}\,$, $\Theta_{11}\,$, $\Theta_{12}\,$,
$\Theta_{21}\,$, $\Theta_{22}$, $\Sigma_{1}\,$, $\Sigma_{2}$ 
and $\Gamma$ are constant matrices (coupling constants) with
number of rows and columns --that we do not need to specify-- determined 
by the range of the internal indices of the fields that they couple. Let us consider, then the following action
\begin{eqnarray}
S & = & \int_{\cal M}
\left[
	d \au^{\dagger}\wedge\bu
+\frac{1}{2}\bu^{\dagger}\wedge \Omega_{11} \bu
+\bu^{\dagger}\wedge \Omega_{12}\bd
+\frac{1}{2}\bd^{\dagger}\wedge\Omega_{22} \bd
+d\fu^{\dagger}\wedge \cu+
\right.
\nonumber\\
& & \hspace{1cm}
+\au^{\dagger}\Theta_{11} \cu
+\au^{\dagger}\wedge\Theta_{12}\cd
+\ad^{\dagger}\wedge\Theta_{21}\cu
+\ad^{\dagger}\wedge\Theta_{22}\cd+\fu^\dagger \Sigma_{1} D+
\nonumber\\
& & \left.
\hspace{1cm}    
+\fd^{\dagger}\Sigma_{2} D
+\Gamma^{\dagger}D\right]
+\int_{{\cal M}}\hat{T}_{\dots}\tilde{S}^{\dots}\label{002}	\quad.
\end{eqnarray}

Notice that we do not need to include additional 
coupling matrices 
in the derivative terms because they can be written as in (\ref{002}) 
by a linear
redefinition and the convention that fields
that do no couple to derivative terms are classified as type 2. 
Specifically, if $\Psi\in{\cal M}_{_{M\times N}}(\real)$ and we have
$dA^{\dagger}\wedge\Psi B$, we can introduce bases
for $\real^{N}$ and $\real^{M}$ as 
${\cal B}_{B}=\{v_{1},\dots,v_{r},\rho_{1},\dots,\rho_{N-r}\}\,$,
${\cal B}_{A}=\{w_{1},\dots,w_{r},\lambda_{1},\dots,\lambda_{M-r}\}\,$
where $r={\rm rank}(\Psi)$, $\Psi \rho_{k}=0$ for
$k=1,\dots,N-r$, $\lambda_{j}^{\dagger}\Psi=0$ 
for $j=1,\dots,M-r$ and the $v$'s and $w$'s
are chosen so that ${\cal B}_{A}$ and ${\cal B}_{B}$ 
are actually bases for the corresponding vector spaces. We have now
\begin{eqnarray}
dA^\dagger\wedge \Psi B=
dA^{\dagger}({\cal B}_{A}^{-1})^{\dagger}
{\cal B}_{A}^{\dagger}\Psi{\cal B}_{B}
{\cal B}_{B}^{-1}B\label{003}\quad,
\end{eqnarray}
where ${\cal B}_{A}^{\dagger}\Psi{\cal B}_{B} $ has the following block form
\begin{eqnarray}
\left[   \begin{array}{cc}
[w_{a}^{\dagger}\Psi v_{b}]   &0\\0&0\end{array}\right]
\label{004}\quad,
\end{eqnarray}
$[w_{a}^{\dagger}\Psi v_{b}] \in{\cal M}_{r\times r}(\real)$ and is regular
so that by independent linear redefinitions in $A$ and $B$ 
it can be taken to be the identity.
By using the convention that fields
that do no couple to derivatives are
``type 2'' we see that the derivative terms can be taken as in (\ref{002})
with all generality and in particular, that the number of internal
components in $\au$ and $\bu$ can be taken to be the same\footnote{A 
similar argument applies to $C$ and $\varphi$.}.

	The last term in the action (\ref{002}) is built out of fields
that do not couple either directly nor indirectly with
the type 1, 2 fields. These fields can be any pair of tensor densities
with matching (covariant and contravariant)
tangent space indices and any kind of internal indices (with the 
corresponding coupling matrices). As we will show in due time these terms 
can be treated separately and it is straightforward to see 
that they do not describe neither local nor topological degrees of
freedom.

	Our purpose now is to study the dynamical content of the
action introduced above, specifically
we want to describe its physical degrees of freedom and
gauge symmetries. A possible line of attack to this problem would be to
use Hamiltonian methods (Dirac analysis of constraints and so on).
This can be done in principle but is very messy in practice. The reason
is that the process of finding secondary constraints is complicated 
by the fact that they can (and do) appear as consistency conditions
in the equations that determine the Lagrange multipliers 
introduced in Dirac's method. As we impose no regularity conditions on the
coupling matrices the process gets quite involved.

		Fortunately there are other methods available
that are specifically suited for the case we are considering here, 
these are the covariant methods considered by Witten and
Crnkovic \cite{EW2}. In essence they consist on working
directly on the solution space to the field equations and build, from
the action, the symplectic structure on this space. If one
has a complete characterization of the solutions depending on 
arbitrary parameters (that may be fields) one can compute the restriction
of the symplectic form to the solution subspace, the gauge directions 
(degenerate directions of the symplectic 
form on the solutions) and identify
the physical degrees of freedom. 
This is usually difficult to achieve for interacting Lagrangians but
can be easily done for the action (2).

\section{Solving the Field Equations}

The field equations derived from (\ref{002}) can be written
in compact form as
\begin{eqnarray}
\Sigma^{\dagger}\varphi+\Gamma^{\dagger}&=&0\label{005}\\
\p\!_{_{\varphi}}d\varphi+\Theta^{\dagger}A&=&0\label{006}\\
 \p\!_{_{A}}dA+\Omega B&=&0\label{007}\\    
 \p\!_{_{B}}dB+\Theta C&=&0\label{008}\\    
 \p\!_{_{C}}dC-\Sigma D&=&0\label{009}
\end{eqnarray}
plus the equations derived from the $\hat{T}_{\dots}\tilde{S}^{\dots}$
terms. We have used the following compact notation
\begin{eqnarray}
\Omega\equiv 
	\left[
	\begin{array}{cc} \Omega_{11}&\Omega_{12}\\
				\Omega_{12}^{\dagger}&\Omega_{22}\end{array}
	\right]\quad;\quad
\Theta\equiv 
	\left[
	\begin{array}{cc} \Theta_{11}&\Theta_{12}\\
				\Theta_{21}&\Theta_{22}\end{array}
	\right]\quad;\quad
\Sigma\equiv 
	\left[
	\begin{array}{c} \Sigma_{1}\\
				\Sigma_{2}\end{array}
	\right]\quad;\quad           \nonumber\\
\varphi\equiv 
	\left[
	\begin{array}{c} \fu\\ \fd\end{array}
	\right]\quad;\quad           
A\equiv 
	\left[
	\begin{array}{c} \au\\ \ad\end{array}
	\right]\quad;\quad           
B\equiv 
	\left[
	\begin{array}{c} \bu\\ \bd\end{array}
	\right]\quad;\quad           
C\equiv 
	\left[
	\begin{array}{c} \cu\\ \cd\end{array}
	\right]\quad,\quad           
\nonumber        
\end{eqnarray}
and   $\p\!_{_{\varphi}}$, $\p\!_{_{A}}$,
$\p\!_{_{B}}$, $\p\!_{_{C}}$ are projectors on the 1 part 
of $\varphi$, $A$, $B$, $C$
\begin{eqnarray}
	\left[
	\begin{array}{c} \fu\\ \fd\end{array}
	\right]            
	\stackrel{\p\!_{_{\varphi}}}{\mapsto}
	\left[
	\begin{array}{c} \fu\\ 0\end{array}
	\right]\quad;\quad           
	\left[
	\begin{array}{c} \au\\ \ad\end{array}
	\right]            
	\stackrel{\p\!_{_{A}}}{\mapsto}
	\left[
	\begin{array}{c} \au\\ 0\end{array}
	\right]\quad;\quad           \nonumber\\
	\left[
	\begin{array}{c} \bu\\ \bd\end{array}
	\right]            
	\stackrel{\p\!_{_{B}}}{\mapsto}
	\left[
	\begin{array}{c} \bu\\ 0\end{array}
	\right]\quad;\quad           
	\left[
	\begin{array}{c} \cu\\ \cd\end{array}
	\right]            
	\stackrel{\p\!_{_{C}}}{\mapsto}
	\left[
	\begin{array}{c} \cu\\ 0\end{array}
	\right]\quad,\quad           \nonumber
\end{eqnarray}        
respectively. The previous equations can be solved in successive 
steps starting from (\ref{005}).

The solution to (\ref{005}) is 
\begin{eqnarray}
\varphi(x)=\lambda_{\sigma}^{\Sigma}\varphi^{\sigma}(x)
	-\Sigma_{_{-1}}^{\dagger}\Gamma^\dagger\label{010}\quad,
\end{eqnarray}
where   $(\lambda_{\sigma}^{\Sigma})^{\dagger}\Sigma=0$ and $\sigma$ labels a 
linearly independent set of left zero-eigenvectors of $\Sigma$,
$-\Sigma_{_{-1}}^{\dagger}\Gamma^\dagger  $ denotes a particular solution
to the equation $\Sigma^{\dagger}\varphi=-\Gamma^\dagger$, 
$\varphi^{\sigma}(x)$ are --at this stage-- arbitrary functions, and
finally, $\Gamma$ must be subject to the consistency condition
$\Gamma\rho^{\Sigma}_{s}=0$ (where $\Sigma\rho_{s}^{\Sigma}=0$ and
$s$ labels a linearly independent set 
of right zero-eigenvectors of $\Sigma$).

  Plugging (\ref{010}) into (\ref{006}) we get 
  \begin{eqnarray}
  \p\!_{_{\varphi}}\lambda^{\Sigma}_{s}d\varphi^{s}(x)
  +\Theta^{\dagger}A(x)=0\label{011}\quad.
  \end{eqnarray}
Equation (\ref{011}) gives the consistency condition 
\begin{eqnarray}
[(\rho^{\Theta}_{c})^{\dagger}\p\!_{_{\varphi}}\lambda^{\Sigma}_{\sigma}]
d\varphi^{\sigma}(x)\equiv {\cal M}_{c\sigma}d\varphi^{\sigma}(x)=0\label{012}\quad,
\end{eqnarray}
(where $\Theta\rho^{\Theta}_{c}=0$ and $c$ labels right zero-eigenvectors)  
and allows us to solve for $A(x)$ from (\ref{011})
when (\ref{012}) holds 
\begin{eqnarray}
A(x)=-[\Theta_{_{-1}}^{\dagger}\p\!_{_{\varphi}}\lambda^{\Sigma}_{\sigma}]
d\varphi^{\sigma}(x)+\lambda_{\theta}^{\Theta}A^{\theta}(x)\label{013}
\quad.
\end{eqnarray}
In analogy with the previous steps the first term of the r.h.s.
of (\ref{013}) is just a particular solution to (\ref{011}) and $\lambda_{\theta}^{\Theta}$
satisfy $(\lambda_{\theta}^{\Theta})^{\dagger}\Theta=0$. 
$A^{\theta}(x)$ are, at the moment, arbitrary 1-forms. Equation (\ref{013})
will be used in the next step of process of solving the equations. To 
complete this step we need to solve (\ref{012}). The dimensions 
of the matrix ${\cal M}_{c\sigma}$ are determined by the dimensions
of ${\rm ker}_{_{R}}\Theta$ and  ${\rm ker}_{_{L}}\Sigma$. To solve
(\ref{012}) we expand 
\begin{eqnarray}
\varphi^{\sigma}(x)=\varphi^{p_{_{0}}}(x)[\rho^{{\cal M}}_{p_{_{0}}}]^{\sigma}
			+
\varphi^{q_{_{0}}}(x)[v^{{\cal M}}_{q_{_{0}}}]^{\sigma}\label{014}\quad,
\end{eqnarray}
where  $[\rho^{{\cal M}}_{p_{_{0}}}]^{\sigma} $, $p_{_{0}}=1,\dots,{\rm 
dim \, ker}_{_{R}}{\cal M}$ 
are a complete set of right zero-eigenvectors labeled by $p_{_{0}}\,$;
here $\sigma$ explicitly labels the rows of each of these vectors. 
The  $[v^{{\cal M}}_{q_{_{0}}}]^{\sigma} $  are vectors
in the  orthogonal complement of ${\rm ker}_R {\cal M}$ 
labeled by $q_0\,$. 
Together with $\rho^{{\cal M}}_{p_0} $ they form
a basis of $\real^{{\rm dim\,ker}_L\Sigma}$.

Introducing (\ref{014}) in (\ref{012}) we are left with the equation
\begin{eqnarray}
{\cal M}_{c\sigma}[ v^{{\cal M}}_{q_{_{0}}}]^{\sigma} d\varphi^{q_{_{0}}}(x)=0
\label{015}             \quad.
\end{eqnarray}
As the restriction of ${\cal M}_{c\sigma}$ to the vector subspace 
generated by $\left\{ v^{{\cal M}}_{q_{_{0}}}\right\} $ is non-singular the 
previous equation implies 
$d\varphi^{q_{_{0}}}(x)=0$ and hence
\begin{eqnarray}
\varphi^{q_{_{0}}}(x)=f\,^{q_{_{0}}i_{_{0}}}\varphi_{_{i_{_{0}}}}(x)
\label{016}\quad,
\end{eqnarray}
where $f\,^{q_{_{0}}i_{_{0}}}$ are arbitrary real numbers;
$\{\varphi_{_{i_{_{0}}}}\}_{i_0=1}^{{\rm dim}H^0({\cal M})} \subset 
H^{0}({\cal M})$ 
(zero de Rham cohomology group of ${\cal M}$; see appendix 2 for a brief introduction to the de Rham cohomology)  form a basis; 
that is $i_{_{0}} $ goes from
 1 to the number of connected components of $\cal M$. Wrapping up
 the previous results we conclude that
 \begin{eqnarray}
     \varphi(x)=\left\{ 
       f\,^{q_{_{0}}i_{_{0}}}  \varphi_{_{i_{_{0}}}}(x)     
+  \varphi^{p_0}(x)  [\rho^{{\cal M}}_{p_{_{0}}}]^{\sigma}
      \right\}\lambda^{\Sigma}_{\sigma}\label{017}
      \quad.
 \end{eqnarray}
We also have a partial 
solution to $A(x)$ given  given by (\ref{013}) that we need 
in order to continue
with the resolution process. We leave the details for the  appendix 3; 
here we just give the final result
\begin{eqnarray}
\varphi(x)&=&\left\{ 
       f\,^{q_{_{0}}i_{_{0}}}  \varphi_{_{i_{_{0}}}}(x)     
+  \varphi^{p_{_{0}}}(x)  [\rho^{{\cal M}}_{p_{_{0}}}]^{\sigma}
      \right\}\lambda^{\Sigma}_{\sigma}
      -\Sigma_{_{-1}}^\dagger\Gamma^\dagger
      \label{18a}\\
A(x)&=&
	-\left\{
		\Theta_{_{-1}}^{\dagger}\p\!_{_{\varphi}}\lambda^{\Sigma}_{\sigma}
	\right\}
	[\rho^{\cal M}_{p_{_{0}}}]^{\sigma}d \varphi^{p_{_{0}}}(x)
	\nonumber
	\\
	&+&\left\{
		 A^{ p_{_{1}}}(x)[\rho^{\cal N}_{p_{_{1}}} ]^{\theta}
		+
		( \alpha^{q_{_{1}} i_{_{1}}} A_{i_{_{1}}}(x)
		+d \varpi^{q_{_{1}}}_{0}(x))
		[v^{\cal N}_{q_{_{1}} }]^{\theta}
	 \right\} \lambda^{\Theta}_{\theta}\label{18b}
\\
B(x)&=&
	-\left\{
		\Omega_{_{-1}}^{\dagger}\p\!_{_{A}}\lambda^{\Theta}_{\theta}
	\right\}
	[\rho^{\cal N}_{p_{_{1}}}]^{\theta}d A^{p_{_{1}}}(x)
	\nonumber
	\\
	&+&\left\{
		 B^{ p_{_{2}}}(x)[\lambda^{\cal N}_{p_{_{2}}} ]^{w}
		+
		( \beta^{q_{_{2}} i_{_{2}}} B_{i_{_{2}}}(x)
		+d \varpi^{q_{_{2}}}_{1}(x))
		[v^{{\cal N}^{\dagger}}_{q_{_{2}} }]^{w}
	 \right\} \rho^{\Omega}_{w}\label{18c}
\\
C(x)&=&
	-\left\{
		\Theta_{_{-1}}\p\!_{_{B}}\rho^{\Omega}_{w}
	\right\}
	[\lambda^{\cal N}_{p_{_{2}}}]^{w}d B^{p_{_{2}}}(x)
	\nonumber
	\\
	&+&\left\{
		 C^{ p_{_{3}}}(x)[\lambda^{\cal M}_{p_{_{3}}} ]^{c}
		+
		( \gamma^{q_{_{3}} i_{_{3}}} C_{i_{_{3}}}(x)
		+d \varpi^{q_{_{3}}}_{2}(x))
		[v^{\cal M}_{q_{_{3}} }]^{c}
	 \right\} \rho^{\Theta}_{c}\label{18d}
\\
D(x)&=&\left\{ \Sigma_{_{-1}}\p\!_{_{C}}\rho^{\Theta}_{\theta}\right\}
	[\lambda^{\cal M}_{p_{_{3}}}]^{c}d C^{p_{_{3}} }(x)
	+ D^{s}(x)\rho^{\Sigma}_{s}\label{18e}
	\quad.
\end{eqnarray}
In the previous examples the only matrices needed 
are ${\cal M}_{c\sigma}\,$,
${\cal N}_{w\theta}$ 
and their transposes as a consequence
of the fact that
\begin{eqnarray}
{\cal M}_{c\sigma}&=&(\rho^{\Theta}_{c})^{\dagger}\p\!_{_{\varphi}}
	\lambda^{\Sigma}_{\sigma}=(\lambda^{\Lambda}_{\sigma})^{\dagger}
	\p\!_{_{C}}\rho^{\Theta}_{c}={\cal M}^\dagger_{\sigma c}
	\nonumber\\
{\cal N}_{w\theta}&=&(\rho^{\Omega}_{w})^{\dagger}\p\!_{_{A}}
	\lambda^{\Theta}_{\theta}=(\lambda^{\Theta}_{\theta})^{\dagger}
	\p\!_{_{B}}\rho^{\Omega}_{w}={\cal N}^\dagger_{\theta w}
	\quad.\label{19}
\end{eqnarray}
We have used the notation explained in appendix 3. Here we
just discuss the general features of (\ref{18a}--\ref{18e}). First we notice
that the solution is parametrized by three types
of objects: constant parameters 
$f^{q_{_{0}}i_{_{0}}}$, 
$\alpha^{q_{_{1}}i_{_{1}}}$, 
$\beta^{q_{_{2}}i_{_{2}}}$,
$\gamma^{q_{_{3}}i_{_{3}}}$ 
that multiply elements of the bases of cohomology groups 
$H^{0}({\cal M}),\cdots,H^{3}({\cal M})$ and two
sets of differential forms $\varphi^{p_{_{0}}}(x)$,
$A^{p_{_{1}}}(x)$, $B^{p_{_{2}}}(x)$, $C^{p_{_{3}}}(x)$, $D^{s}(x)$ 
and 
$\varpi^{q_{_{1}}}_{0}(x)$,
$\varpi^{q_{_{2}}}_{1}(x)$, $\varpi^{q_{_{3}}}_{2}(x)$ 
(or rather 
$d\varpi^{q_{_{1}}}_{0}(x)$,
$d\varpi^{q_{_{2}}}_{1}(x)$, $d\varpi^{q_{_{3}}}_{2}(x)$ ).
It is important to emphasize 
that we are not entitled to disregard neither 
$A^{p_{_{1}}}(x)$, $B^{p_{_{2}}}(x)$, $C^{p_{_{3}}}(x)$, $D^{s}(x)$ 
nor
$\varpi^{q_{_{1}}}_{0}(x)$,
$\varpi^{q_{_{2}}}_{1}(x)$ and $\varpi^{q_{_{3}}}_{2}(x)$
without knowing that they are gauge parameters; a fact that 
we {\it ignore} at this stage.

\section{The Symplectic Structure}

	Given a quadratic action the problem of finding out its
gauge symmetries is intimately connected with that of solving the
Euler-Lagrange equations. If we formally write a quadratic
action for a set of fields $\{\phi^{i}\}_{i\in I}$
\begin{eqnarray}
S[\phi]=\int \phi^{i}A_{ij}\phi^{j}\label{021}\quad,
\end{eqnarray}
and $A$ a field independent symmetric linear operator, the variation of $S$ 
when we change $\phi^{i}$ by $\phi^{i} +\delta\phi^{i}$ is 
given by the exact expression
\begin{eqnarray}
\delta S=\int[\delta\phi^{i}A_{ij}\phi^{j}
+\phi^{i}A_{ij}\delta\phi^{j}+\delta\phi^{i}A_{ij}\delta\phi^{j}]\label{022}\quad,
\end{eqnarray}
whereas the Euler-Lagrange equations are $A_{ij}\phi^{j}=0$. 
If $\chi^{i}$ is such that $A_{ij}\chi^{j}=0$ and we consider {\it any} 
field configuration $\phi^{i}$ the action is invariant under the transformation 
$\phi^{i}\mapsto\phi^{i}+\chi^{i}$ as can be seen from (\ref{022}). Of course, 
in non-gauge invariant theories,
this is just the linear superposition principle.
A gauge symmetry would manifest itself 
in a similar way but now it should be thought of
as an arbitrariness
in a set of solutions corresponding, loosely speaking,
to the same initial conditions\footnote{One should take into account the 
possibility of using different initial data hypersurfaces.}. The question
is then: How do we discriminate between both situations?

	One could be tempted to think that if we happen
to have a complete parametrization of the solutions to the field 
equations, depending on a certain set of parameters (that can 
actually be functions and should, at least, describe all the possible
initial conditions) it is straightforward to tell gauge parameters apart
from the other quantities needed to specify a solution. It is very
important to understand that it is not the case. If we are given a subset of 
a space field configurations parametrized by
certain functions there is no way to decide, in a 
uniquely consistent manner, which of them are gauge parameters and which 
ones parametrize
inequivalent configurations. In order to do that an extra element
is needed: a suitably defined symplectic form in the space of fields. 
Though it is more frequent to  find the symplectic structure
in the Hamiltonian formulation there are covariant methods \cite{EW2}
that allow us to work directly with field configurations.
The necessity of a symplectic form is clear in the Hamiltonian 
framework. Given a phase space and certain submanifold
of it (let us suppose that the Hamiltonian is zero)
we need the symplectic form $\omega$  to define
what we understand as gauge orbits and gauge transformations
(this is done by looking at degenerate directions of $\omega$ on 
the constraint hypersurface). The situation is analogous in the 
covariant setting.

\bigskip

	In sight of (\ref{18a}--\ref{18e}) one is tempted to conclude that the action (\ref{002}) just describes the topological degrees of freedom labeled by 
$f^{q_{_{0}}i_{_{0}}}$, 
$\alpha^{q_{_{1}}i_{_{1}}}$, 
$\beta^{q_{_{2}}i_{_{2}}}$,
$\gamma^{q_{_{3}}i_{_{3}}}\,$. 
In simple examples it is possible, in practice, 
to directly write down  a set of independent gauge transformations
and --actually guess-- the physical degrees of freedom. 
This is not that  easy 
here and a different method must be used. In our case, 
and following \cite{EW2},
the symplectic structure is given by
\begin{eqnarray}
\omega=\int_{\Sigma}J
=\int_{\Sigma}[{\rm d}\! {\rm I}\au^{\dagger}\wedge\!\!\!\!\!\!\wedge 
\,{\rm d}\! {\rm I}\bu+{\rm d}\! {\rm I}\fu^{\dagger}
\wedge\!\!\!\!\!\!\wedge \,{\rm d}\!{\rm I}\cu]\quad.\label{023}
\end{eqnarray}
In the previous expressions ${\rm d}\!{\rm I}$ and $ 	\wedge \!\!\!\!\!\wedge$ 
denote the exterior differential and product
in the field space {\it coordinatized} by $\varphi(x)$, $A(x)$, $B(x)$, 
$C(x)$ and $D(x)$. The exterior 
product in $\cal M$ is implicitly understood in (\ref{023}).
 In order to perform the three dimensional integral
appearing in (\ref{023}) one must specify  the three 
manifold $\Sigma$. That the result is independent of $\Sigma$ is
a consequence of the fact that $dJ=0$ on solutions to the 
field equations (here $d$ is the exterior differential on $\cal M$).
This can be checked in a straightforward way by acting
on $J$ with
$d$; keeping track of the form order both in $\cal M$ 
and in the field space (to get the minus signs right) and using the field equations
(\ref{005}--\ref{009}) after explicitly splitting them in type 
1 and type 2 fields.

	Once we have $\omega$ we must compute it on the
solutions to the fields equations given by (\ref{18a}--\ref{18e}), 
as a previous step
to finding the degenerate directions. This way we get
\begin{eqnarray}
\left. \omega \right|_{sol}&=&
	[ v^{\cal M}_{ q_{_{0}} }]^{\sigma}
	{\cal M}^{\dagger}_{\sigma c}  
	[ v^{ {\cal M}^{\dagger} }_{ q_{_{3}} }]^{c}   
	\left( \int_{\Sigma} \phi_{ i_{_{ 0 }} }C_{i_{_{3}}} \right)
	{\rm d}\! {\rm I} f^{ q_{_{0}}i_{_{ 0 }} } 
	\wedge\!\!\!\!\!\!\wedge\,{\rm d}\! {\rm I}\gamma^{q_{_{3}}i_{_{3}}}
	\nonumber
	\\
	&+&
	[v^{\cal N}_{q_{_{1}}}]^{\theta}
	{\cal N}^{\dagger}_{\theta w}  
	[v^{{\cal N}^{\dagger}}_{q_{_{2}}}]^{w}   
	\left(\int_{\Sigma} A_{ i_{_{1}} }\wedge B_{i_{_{2}}}\right)
	{\rm d}\! {\rm I} \alpha^{q_{_{1}}i_{_{1}}} 
	\wedge \!\!\!\!\!\!\wedge\,{\rm d}\! {\rm I} 
	\beta^{q_{_{2}}i_{_{2}}}\label{024}\quad.
\end{eqnarray}
Several points are now in order. First of all the fact that the 
only parameters that appear in  $\left. \omega \right|_{sol}$
are 
$f^{q_{_{0}}i_{_{0}}}$, 
$\alpha^{q_{_{1}}i_{_{1}}}$, 
$\beta^{q_{_{2}}i_{_{2}}}$ and
$\gamma^{q_{_{3}}i_{_{3}}}$ 
shows that the theory has no local degrees of freedom 
( $\left. \omega \right|_{sol}$  acting on tangent vectors  to the
solution manifold is zero whenever these vectors 
``point in the directions'' parametrized by all the $x$-dependent
fields). Second, even though one would naively 
expect that the only parameters appearing 
in (\ref{024}) are those corresponding to the ``topological
sectors'' the fact that $x$-dependent fields do not appear in (\ref{024}) depends
on non-trivial cancellations between terms coming
from the first wedge product in (\ref{023}) and terms coming from the second
(see appendix 4).  Third, from the K\"unneth formula and the fact that
the cohomology groups of $\real$ are $H^0(\real)=\real$ and $H^1(\real)=0$
we conclude that $H^s({\cal M})=H^s(\Sigma)$, in 
particular notice that this is the reason why $H^4({\cal M})$ 
does not appear.
Finally, each of the two terms in (\ref{024}) has a coefficient (a matrix in the 
pairs of indices labeling $f$, $\alpha$, $\beta$, $\gamma$)
that could, in principle, be degenerate. A fact that would imply the 
presence of additional gauge transformations involving now the topological
sectors. This, however, is not the case. On one hand 
both integrals
\begin{eqnarray}
\int_{\Sigma}  \varphi_{i_0 }\,
		C_{i_3}\quad;\quad
\int_{\Sigma} A_{i_1}\wedge B_{i_2} \label{025}
\end{eqnarray}
are non singular square matrices (see appendix 2).
On the other both
\begin{eqnarray}
	[ v^{\cal M}_{ q_{_{0}} }]^{\sigma}
	{\cal M}^{\dagger}_{\sigma c}  
	[ v^{ {\cal M}^{\dagger} }_{ q_{_{3}} }]^{c}   
	\quad,\quad
	[v^{\cal N}_{q_{_{1}}}]^{\theta}
	{\cal N}^{\dagger}_{\theta w}  
	[v^{{\cal N}^{\dagger}}_{q_{_{2}}}]^{w}   
\end{eqnarray}
are easily proved to be square non-singular matrices so that we 
finally conclude that (\ref{024}) contains no degenerate directions.
Taking appropriate bases in the cohomology groups it is 
straightforward to write (\ref{024}) in canonical form.

	At this point it is a simple matter to deal with  the
additional $\hat{T}_{\dots}\tilde{S}^{\dots}$ terms in the original action.
As they do not involve derivatives they do not appear in the symplectic 
structure (and its restriction to the solution subspace)
which means that all the degrees of freedom  described by them are pure 
gauge.

\section{Degrees of Freedom and Gauge Transformations}

	We summarize our previous results from the last sections. 
	As we have just seen the only physical degrees of freedom described
	by (\ref{002}) are purely topological and described
	by pairs of variables 
	$(f^{q_{_{0}} i_{_{0}}}, \gamma^{q_{_{3}} i_{_{3}}})\,$
	and
	$(\alpha^{q_{_{1}} i_{_{1}}}, \beta^{q_{_{2}} i_{_{2}}})\,$.
   The gauge symmetries of the action (\ref{002})
are
\begin{eqnarray}
\delta \varphi(x)&=& 
  \delta \varphi^{p_{_{0}}}(x)  [\rho^{{\cal M}}_{p_{_{0}}}]^{\sigma}
	\lambda^{\Sigma}_{\sigma}\label{026}\\
\delta A(x)&=&
	-\left\{
		\Theta_{_{-1}}^{\dagger}\p\!_{_{\varphi}}\lambda^{\Sigma}_{\sigma}
	\right\}
	[\rho^{\cal M}_{p_{_{0}}}]^{\sigma}d \delta  \varphi^{p_{_{0}}}(x)
	\nonumber
	\\
	&+&\left\{
		\delta  A^{ p_{_{1}}}(x)[\rho^{\cal N}_{p_{_{1}}} ]^{\theta}
		+
		d  \delta \varpi^{q_{_{1}}}_{0}(x)
		[v^{\cal N}_{q_{_{1}} }]^{\theta}
	 \right\} \lambda^{\Theta}_{\theta}\nonumber
\\
\delta B(x)&=&
	-\left\{
		\Omega_{_{-1}}^{\dagger}\p\!_{_{A}}\lambda^{\Theta}_{\theta}
	\right\}
	[\rho^{\cal N}_{p_{_{1}}}]^{\theta}d \delta  A^{p_{_{1}}}(x)
	\nonumber
	\\
	&+&\left\{
		 \delta B^{ p_{_{2}}}(x)[\lambda^{\cal N}_{p_{_{2}}} ]^{w}
		+
		d \delta \varpi^{q_{_{2}}}_{1}(x)
		[v^{{\cal N}^{\dagger}}_{q_{_{2}} }]^{w}
	 \right\} \rho^{\Omega}_{w}\nonumber
\\
\delta C(x)&=&
	-\left\{
		\Theta_{_{-1}}\p\!_{_{B}}\rho^{\Omega}_{w}
	\right\}
	[\lambda^{\cal N}_{p_{_{2}}}]^{w}d \delta B^{p_{_{2}}}(x)
	\nonumber
	\\
	&+&\left\{
		 \delta C^{ p_{_{3}}}(x)[\lambda^{\cal M}_{p_{_{3}}} ]^{c}
		+d \delta \varpi^{q_{_{3}}}_{2}(x)
		[v^{\cal M}_{q_{_{3}} }]^{c}
	 \right\} \rho^{\Theta}_{c}\nonumber
\\
\delta D(x)&=&\left\{ \Sigma_{_{-1}}\p\!_{_{C}}\rho^{\Theta}_{\theta}\right\}
	[\lambda^{\cal M}_{p_{_{3}}}]^{c}d\delta C^{p_{_{3}} }(x)
	+\delta D^{s}(x)\rho^{\Sigma}_{s}\nonumber\quad,
\end{eqnarray}
where the gauge parameters are 
the 0-forms $\delta \varphi^{p_{_{0}}}(x)$,$\delta\varpi^{q_{_{1}}}_{0}(x)$, ‡
the 1-forms $ \delta  A^{ p_{_{1}}}(x)$, $\delta \varpi^{q_{_{2}}}_{1}(x)$, 
the 2-forms $\delta B^{p_2}(x)$, $\delta \varpi^{q_3}_2(x)$,   
the 3-forms $\delta C^{ p_{_{3}}}(x)$ 
and the 4-forms $\delta D^{s}(x)$. 
   Though, {\it a posteriori}, the structure of these transformations 
is quite simple it is not straightforward to guess the complicated 
matrix coefficients appearing in the previous expressions.

 We consider now some examples of these types of theories that have
been studied in the literature \cite{HB}, \cite{FER}. In them some of the 
fields ($\varphi$ and $C$) do not appear. This fact is taken into account
 in (\ref{026}) by putting $P_{_\varphi}=P_{_C}=0$ and realizing 
 that $\delta \varphi$ and $\delta C$ are then arbitrary
 and do not affect both $\delta A$ and $\delta B$.
\begin{itemize}

 \item{\bf Example 1:} Abelian $BF$ model with no internal indices. Here
 \begin{eqnarray}
 S=\int_{\cal M} dA\wedge B\quad,
 \end{eqnarray}
so that $\Omega=0$, and the other fields do not appear. In this case
${\cal M}_{c\sigma}=0$, ${\cal N}_{\theta w}=1$;
$v^{\cal N}=1$, $v^{\cal N^{\dagger}}=1$, then
$\left.\omega\right|_{sol}
=\left(\int_\Sigma A_{i_1}\wedge B_{i_2}\right) 
	{\rm d\! I}\alpha^{i_1} 
	\wedge \!\!\!\!\!\wedge
	{\rm d\! I}\beta^{i_2}$
and the theory has ${\rm dim} \, H^1(\Sigma)
=    {\rm dim} \, H^2(\Sigma)
$ degrees of freedom.

	The gauge transformations are
\begin{eqnarray}
\delta A(x)&=&d \delta \varpi_{0}(x)\\
\delta B(x)&=&d \delta \varpi_{1}(x)\quad.\nonumber
\end{eqnarray}

\item {\bf Example 2:} 
\begin{eqnarray}
S=\int_{\cal M}[dA\wedge B -\frac{1}{2}B\wedge B]\quad,
\end{eqnarray}
so that 
$\Omega=-1$, 
 ${\cal M}_{c\sigma}=0$, 
${\cal N}_{\theta w}=0$;
$v^{\cal N}=v^{\cal M}=0$.
This means that, on solutions, 
$\left.\omega \right|_{sol}= 0$ so that all the points in the 
solution subspace are in the same gauge orbit and hence the theory
has zero degrees of freedom. The gauge transformations are 
\begin{eqnarray}
\delta A(x)&=&\delta \chi(x)\\
\delta B(x)&=&d \delta \chi(x)\nonumber      \quad.
\end{eqnarray}

\item {\bf Example 3:}\footnote{This example appears as the quadratic 
term in the formulation of the Husain-Kucha{\v r} model as a coupled
$BF$ system \cite{FER}.}
\begin{eqnarray}
S=\int_{\cal M}[dA_1\wedge B_1+dA_2\wedge B_2+B_1\wedge B_2]\quad,
\end{eqnarray}
with 
$\Omega={0\,1\choose 1\,0}$,  
     $\p\!_{_A}={1\,0\choose 0\,1}$,
 $\p\!_{_B}= {1\,0\choose 0\,1}$, 
 $\rho^\Omega=\left[{0\atop 0}\right]$, 
  ${\cal M}_{c\sigma}=0$,
   ${\cal N}_{w\theta}=0$,
   so that $v^{\cal N}=v^{\cal M}=0$, 
    $\rho^{\cal N}_1=\left[{1\atop 0}\right]$ 
    and $\rho^{\cal N}_2=\left[{0\atop 1}\right]\,$,
and, as before, the theory has zero degrees of freedom. 
The gauge transformations are
\begin{eqnarray}
\delta A_1(x)&=&\delta \chi_1(x)\nonumber\\
\delta A_2(x)&=&\delta \chi_2(x)\\
\delta B_1(x)&=&-d\delta \chi_2(x)\nonumber\\
\delta B_2(x)&=&-d\delta \chi_1(x)\nonumber\quad.
\end{eqnarray}
Notice how the switch in the roles of $\delta \chi_1(x)$ and 
$\delta \chi_2(x)$ in the gauge transformations for the $B_1(x)$, 
$B_2(x)$ fields is correctly described by the general form 
of the gauge transformations given by 
$\delta B(x)$ in (\ref{025}).

\end{itemize}

\section{Conclusions and Comments}

At this point, and after realizing that the final phase space is that 
of a set of uncoupled $BF$ and $C\varphi$ theories,
the attentive reader may wonder if it would not be easier to simplify 
the action from the start by solving for the 
purely algebraic fields equations and
substituting the solution back in the action
in order to eliminate, for example, the type 2 fields and work with a 
simple set of $BF$  and $C\varphi$ theories. Thought, in actual examples,  this may be
a useful strategy there is a catch. The algebraic field equations
usually allow to solve only for combinations of type 1 and type 2 fields
so that, in the end, one may not find a much simpler action. That 
this is so is actually proved by the
rather tangled nature of the gauge transformations (\ref{026}) and specially
by the  awkward matrices that appear in the solution.
A  different manifestation of the same
problem appears if one tries to work within 
the Hamiltonian framework.

	The type of theory described by the action considered above
may be taken as the building block of more
complicated topological and non--topological actions obtained from
(\ref{002})  by adding higher power
terms to it. Their symmetries must be understood,
among other things, in order to treat them perturbatively
(see, for example, \cite{M} and references therein). This is an additional reason to work with an action of the general form introduced above, because it is not 
clear that a non-linear extension of it can be equally 
obtained from a simplified action.

It is straightforward to extend the previous analysis 
to arbitrary dimensions and arrive at the conclusion that
local diff-invariant quadratic Lagrangians in arbitrary space-time 
dimensions do not have local degrees of freedom. This means that 
in order to describe local degrees of freedom with local diff-invariant 
Lagrangians they must be taken to be, at least, cubic in the fields. 
This happens, for example, if the background metric used to write down 
ordinary free actions is taken as a dynamical object. 

	It cannot be overemphasized that without the
use of the symplectic structure there is no way of telling 
which  one of the parameters and functions 
appearing in the general solution to the field equations are gauge
and which ones label genuine degrees of freedom. We feel that the conclusion
that no local degrees of freedom survive is not obvious
{\it a priori}. An important consequence of this result
is that there are no diff-invariant ``lower derivative'' modifications
of the action (\ref{001}) that can be treated perturbatively. Notice, 
however, that our result says nothing against the existence of 
a similar possibility in Palatini-like actions where 
quadratic and cubic terms are not separately
diff-invariant.

	Another interesting consequence of the previous analysis
is related to the meaning of the particle concept in 
diff-invariant theories. In ordinary quantum field theories in a 
Minkowskian background the study of particle states is done 
by first choosing a quadratic (``free'') Lagrangian, finding its 
normal mode expansion 
and labeling the resulting ``elementary excitations'' with 
momentum and spin indices (coming from the Casimir operators
of the space-time symmetry group $ISO(1,3)$). If one 
has a quadratic theory with local degrees of freedom 
in a non Minkowskian background (as one does in QFT's in curved
backgrounds) one still has the normal mode decomposition
but lacks the Poincar\'e symmetry. In this case, though the 
particle interpretation 
is subtler (coordinate dependence, Unruh effect and so on) there is still
one available. What we have found here is that in the absence 
of a background there is no possible particle interpretation.
This result lends support to some approaches to dealing with
diff-invariant QFTs such as the loop variables approach to 
quantum gravity
championed by Ashtekar, Rovelli, Smolin and others \cite{ASH}, \cite{LOOP},
and in particular helps understand why the elementary
excitations (spin network--states and so on) are so different from
the usual Fock-space particle states. 
It may also be worthwhile to point out that string 
theory and its multiple derivatives require the presence of a 
non-dynamical background.

	Finally one may wonder how it would be possible to 
escape the negative conclusions of this paper. If one
is willing to keep diffeomorphism invariance and the quadratic 
character of the Lagrangian one would be forced to abandon locality. If a 
non-local, diff-invariant quadratic action describing propagating
degrees of freedom exists is an open problem that may
be worth thinking about.

\section*{Acknowledgments}

	The authors of the paper 
	wish to thank V. Husain, C. Rovelli, C. Torre, and M. Varadarajan, 
	for comments.

\section*{Appendix 1: Inclusion of Boundaries}

	The result presented in the paper that quadratic, local diff-invariant theories, have no local degrees of freedom has been derived in the case of working with manifolds without a boundary. The attentive reader may  wonder if the result still holds in the presence of boundaries. The answer is in the affirmative as justified in what follows. If $\cal M$ has boundaries the action (\ref{002}) can be generalized by adding the most general diff-invariant action on its boundary $\partial \cal M$
\begin{eqnarray}
S_\partial &=&\int_{\partial {\cal M}}\left[
d\Au^\dagger\wedge\Au+d\Fu^\dagger\wedge\Bu+\Au^\dagger\wedge\theta_{11}\Bu
+\Au^\dagger\wedge\theta_{12}\Bd
+\Ad^\dagger\wedge\theta_{21}\Bu\right.\nonumber\\
&+&\left.\Ad^\dagger\wedge\theta_{22}\Bd
+ \Fu^\dagger\sigma_1c+\Fd^\dagger\sigma_2c+\gamma^\dagger c \right]\label{b}\quad,
\end{eqnarray}
where $\Fu$, $\Fd$ are 0-forms, $\Au$, $\Ad$ are 1-forms, $\Bu$, $\Bd$ are 2-forms and $c$ is a 3-form. One can think of the terms in $S_\partial$ as having two different origins; some of them come from the restriction of the 4-dimensional fields to $\partial \cal M$ whereas others are genuine 3-dimensional fields. If one adopts this point of view the field equations are obtained by varying first in the 4-dimensional fields with variations restricted to be zero on the boundary and then varying the 3-dimensional fields. Another  possible, and equivalent, point of view would be to consider that a part of $S_\partial$ provides the surface terms needed to cancel the surface terms appearing when we vary the 4-dimensional fields and integrate by parts whereas the rest of $S_\partial$ consists of the genuine 3-dimensional objects. In either case we are entitled to treat 4-dimensional and 3-dimensional fields independently, so that in order to prove that no local degrees of freedom appear when boundaries are present it suffices to show that the action (\ref{b}) (defined on the boundary of $\cal M$, $\partial {\cal M}=\real\times \partial\Sigma$, with $\partial^2 \Sigma=\emptyset$ as a consequence of the identity $\partial^2=0$ for the compact 2-manifold $\partial \Sigma$) describes no local degrees of freedom. This is done by following exactly the same steps presented in the main body of the paper for the 4-dimensional case so we do not give the details here.

\section*{Appendix 2: De Rham Cohomology Groups }

	For the benefit of the physics-oriented reader we summarize here the main results about differential forms and the de Rham cohomological groups.

	A manifold $\cal M$ without (with) boundary  of dimension $m$ is a locally $m$-Euclidean (semi $m$-Euclidean) topological space. Smooth manifolds are those that are locally Euclidean in a smooth way. 

	All smooth manifolds (with or without boundary) admit smooth tensor fields. In particular, the completely antisymmetric  $r$-covariant tensor fields are known as $r$-differential forms. Forms can be differentiated by means of the exterior differential operator $d$, that maps $r$-forms in $(r+1)$-forms. The $d$ operator has the property $d^2=0$, that allows to define the cohomology groups on $\cal M$. A $r$-form $w$ on $\cal M$ is closed if $dw=0$ and is exact if $w=dv$ for some $(r-1)$-form $v$ on $\cal M$. Exact forms are all closed because $d^2=0$. The converse is not true as one can see if one defines  an equivalence relation on the vector space of closed $r$-forms: two closed forms $w_1$ and $w_2$ are called cohomologous if $w_1-w_2$ is exact. The set of equivalence classes is denoted by $H^r({\cal M})$, the $r$th de Rham cohomology group of ${\cal M}$. Although  the cohomology groups are defined in terms of the  manifold structure of ${\cal M}$, they are topological invariants; that is, two topologically equivalent manifolds have the same de Rham groups. For compact manifolds they are finite dimensional real vector spaces. Another property that we use in the paper is the fact that for compact manifolds $H^r$ can be identified with $H^{(m-r)}$, this is known as Poincar\'e duality and is a consequence of the fact that integrals of the type given by (\ref{025}) define a non-singular bilinear form in $H^r \times H^{(m-r)}$ \cite{N}. 

Directly from the most general version of the Poincar\'e Lema \cite{G}, that states that  $H^r(\real\times{\cal M})=H^r({\cal M})$, or indirectly, by means of the K\"unneth  formula \cite{N} that relates the cohomology groups of ${\cal M}={\cal M}_1\times{\cal M}_2$ with the ${\cal M}_1\,$, ${\cal M}_2$ ones according to 
\begin{eqnarray*}
H^r({\cal M})=\oplus_{p+q=r}\left( H^p({\cal M}_1)\otimes H^q({\cal M}_2)\right)
\end{eqnarray*}
one can prove $H^r(\real\times \Sigma)=H^r(\Sigma)$, an assertion that we use  in section IV.

\section*{Appendix 3: Solving the Field Equations}

We continue here with the resolution of the
field equations started in section III. Introducing (\ref{013}) in 
(\ref{007}) we get 
\begin{eqnarray}
[\p\!_{_A}\lambda^{\Theta}_{\theta}]dA^\theta(x)+\Omega B(x)=0
\label{A.1}
\end{eqnarray}
which gives, as in the previous step, a consistency condition
and a solution for $B$. The consistency condition is
\begin{eqnarray}
\left\{(\rho^{\Theta}_w)^\dagger\p\!_{_A}\lambda^{\Theta}_{\theta}\right\}
dA^\theta(x)\equiv {\cal N}_{w\theta}dA^\theta(x)=0
\label{A.2}
\end{eqnarray}
(where $\Omega\rho^{\Omega}_{w}=0$ and $w$ labels these 
right zero-eigenvectors) and we get for $B(x)$
\begin{eqnarray}
B(x)=-\left\{\Omega_{_{-1}}\p\!_{_A}\lambda^{\Theta}_{\theta}\right\}
		dA^\theta(x)+B^w(x)\rho^\Omega_w
		\label{A.3}
\end{eqnarray}
The first term in the previous expression is a
particular solution to $(\ref{A.1})$. Expanding 
\begin{eqnarray}
A^\theta(x)=A^{p_1}(x)[\rho^{\cal N}_{p_1}]^\theta
	 +   A^{q_1}(x)[v^{\cal N}_{q_1}]^\theta   
	 \label{A.4}
\end{eqnarray}
with $[\rho^{\cal N}_{p_1}]^\theta$, 
$p_1=1,\dots,{\rm dim\, ker}_{_R}{\cal N}$  a complete generating set 
of right  zero-eigenvectors labeled by $p_1$ ($\theta$ labels the rows 
of each of these vectors) and $[v^{\cal N}_{q_1}]^\theta$
a basis in the complement  of ${\rm ker}_{_R}{\cal N}$.
As before we have to solve the equation
\begin{eqnarray}
dA^{q_1}(x)=0       \label{A.5}
\end{eqnarray}
which gives
\begin{eqnarray}
A^{q_1}(x)=\alpha^{q_1 i_1}A_{i_1}(x)+d\varpi^{q_1}_{0}(x)
\label{A.6}
\end{eqnarray}
Notice that each $A^{q_1}(x)$ is a closed 1-form (an element 
of $Z^1({\cal M})$) that we write as an element in $H^1({\cal M})$ 
plus an arbitrary exact 1-form $d\varpi^{q_1}_0(x)$ 
(where $\varpi^{q_1}_0$ are 0-forms).  The $A_{i_1}(x)$ are a basis
of elements in $H^1({\cal M})$. Plugging $A$ back in (\ref{013}) gives 
(\ref{18b}).

	In the next step we introduce  (\ref{A.3}) in 
(\ref{008}) to get
\begin{eqnarray}
[\p\!_{_B}\rho^{\Omega}_{w}]dB^{w}(x)+\Theta C(x)=0\quad.\label{A.7}
\end{eqnarray}
The consistence condition derived from (\ref{A.7}) is
\begin{eqnarray}
\left\{(\lambda^\Theta_\theta)^\dagger 
	\p\!_{_B}\rho^{\Omega}_{w}\right\}dB^{w}(x)
	\equiv {\cal N}^\dagger_{\theta w}dB^{w}(x)     =0
	\label{A.8}
\end{eqnarray}
and
\begin{eqnarray}
C(x)=-\left\{\Theta_{_{-1}} \p\!_{_B}\rho^{\Omega}_{w}\right\}dB^{w}(x)
	+C^c(x)\rho^\Theta_c\label{A.9}
\end{eqnarray}
the first term in (\ref{A.9})  being a particular solution to 
(\ref{A.7}) and $\Theta\rho^{\Theta}_{c}=0$. Expanding 
\begin{eqnarray}
B^w(x)=B^{p_2}(x)[\lambda^{\cal N}_{p_2}]^w+B^{q_2}(x)[v^{\cal N}_{q_2}]^w
\label{A.10}
\end{eqnarray}
in complete analogy with the previous steps, 
we obtain the equation $dB^{q_2}(x)=0$
\begin{eqnarray}
B^{q_2}(x)=\beta^{q_2 i_2}B_{i_2}(x)+d\varpi^{q_2}_1(x)\label{A.11}
\end{eqnarray}
where $\{B_{i_2}\}$ is a basis of $H^2({\cal M})$ and 
$\varpi^{q_2}_1$ are 1-forms.
Introducing (\ref{A.11}) back in (\ref{A.3}) we 
finally obtain (\ref{18c}).
In the following step we introduce (\ref{A.9}) in (\ref{009}) and obtain 
\begin{eqnarray}
[\p\!_{_C}\rho^{\Omega}_{c}]dC^c(x)-\Sigma D(x)=0              \label{A.12}
\end{eqnarray}
which gives the consistency condition 
\begin{eqnarray}
\left\{(\lambda^\Sigma_\sigma)^\dagger\p\!_{_C}\rho^{\Omega}_{c}\right\}
dC^c(x)=0\label{A.13}
\end{eqnarray}
and
\begin{eqnarray}
D(x)=\left\{\Sigma_{_{-1}} \p\!_{_C}\rho^{\Omega}_{c}\right\}dC^c(x)
+D^s(x)\rho^{\Sigma}_{s}\label{A.13bis}
\end{eqnarray}
As in previous cases 
$ \left\{\Sigma_{_{-1}} \p\!_{_C}\rho^{\Omega}_{c}\right\}dC^c(x)  $ is a 
particular solution to (\ref{A.12}) (as an equation in $D$).
Expanding now
\begin{eqnarray}
C^c(x)=C^{p_3}(x)[\lambda^{\cal M}_{p_3}]^c
	+C^{q_3}(x)[v^{\cal M^\dagger}_{q_3}]^c
	\label{A.14}
\end{eqnarray}
we need to solve $dC^{q_3}(x)=0$ and hence
\begin{eqnarray}
C^{q_3}(x)=\gamma^{q_3 i_3}C_{i_3}(x)+d\varpi^{q_3}_2(x)\label{A.15}
\end{eqnarray}
where the $C_{i_3}(x)$ spans  a basis in $H^3({\cal M})$
and $\varpi^{q_3}_2$ are 2-forms. (\ref{A.15}) allows us to 
obtain (\ref{18d}) from (\ref{A.9}) and (\ref{18e}) from (\ref{A.13bis}).

\section*{Appendix 4: The Symplectic Structure on Solutions}

In order to appreciate the non-triviality of the cancellations
leading to the final symplectic structure (\ref{024}) we 
separately compute
\begin{eqnarray}
\int_\Sigma{\rm d}\! {\rm I}\fu^{\dagger}
\wedge\!\!\!\!\!\!\wedge 
\,{\rm d}\!{\rm I}\cu
&=&
	[ v^{\cal M}_{ q_{_{0}} }]^{\sigma}
	{\cal M}^{\dagger}_{\sigma c}  
	[ v^{ {\cal M}^{\dagger} }_{ q_{_{3}} }]^{c}   
	\left( \int_{\Sigma} \phi_{ i_{_{ 0 }} }C_{i_{_{3}}} \right)
	{\rm d}\! {\rm I} f^{ q_{_{0}}i_{_{ 0 }} } 
	\wedge\!\!\!\!\!\!\wedge\,{\rm d}\! {\rm I}\gamma^{q_{_{3}}i_{_{3}}}
	\nonumber\\
	&-&
\left\{(\lambda^{\Sigma}_{\sigma})^{\dagger}\p\!_{_C}\Theta_{_{-1}}
 \p\!_{_B}\rho^{\Omega}_{w}\right\}[\rho^{\cal M}_{q_0}]^{\sigma}
 \int_\Sigma  {\rm d}\! {\rm I}\varphi^{q_0}
  \wedge\!\!\!\!\!\!\wedge\,
 d{\rm d}\! {\rm I}B^{w}
 \label{A.16}
\end{eqnarray}
with $B^w(x)$ given by (\ref{A.10}) and (\ref{A.11}). 
Notice that, in general,
the last term in the r.h.s. of the  previous expression is not 
necessarily zero.
\begin{eqnarray}
 \int_{\Sigma}{\rm d}\! {\rm I}\au^{\dagger}
 \wedge\!\!\!\!\!\!\wedge 
 \,{\rm d}\! {\rm I}\bu
& =&
	[v^{\cal N}_{q_{_{1}}}]^{\theta}
	{\cal N}^{\dagger}_{\theta w}  
	[v^{{\cal N}^{\dagger}}_{q_{_{2}}}]^{w}   
	\left(\int_{\Sigma} A_{ i_{_{1}} }\wedge B_{i_{_{2}}}\right)
	{\rm d}\! {\rm I} \alpha^{q_{_{1}}i_{_{1}}} 
	\wedge \!\!\!\!\!\!\wedge\,{\rm d}\! {\rm I} \beta^{q_{_{2}}i_{_{2}}}
	\nonumber
	\\
	&+&
\left\{(\lambda^{\Sigma}_{\sigma})^{\dagger}\p\!_{_C}\Theta_{_{-1}}
 \p\!_{_B}\rho^{\Omega}_{w}\right\}[\rho^{\cal M}_{q_0}]^{\sigma}
 \int_\Sigma  {\rm d}\! {\rm I}\varphi^{q_0}
  \wedge\!\!\!\!\!\!\wedge\,
 d{\rm d}\! {\rm I}B^{w}
 \label{A.17}
\end{eqnarray}
after manipulating the matrices appearing in the last term 
of (\ref{A.17}). In the process of obtaining (\ref{A.16}) 
and (\ref{A.17}) several terms 
directly cancel by any of these reasons:
\begin{itemize}
\item[ i.] Appearance of ${\cal M}\rho^{\cal M}_{p_0}\,$.
\item[ii.] $d^{2}=0$.
\item[ iii.] Integration by parts (remember that we take 
$\partial \Sigma =\emptyset$).
\item[iv.] $d\varphi_{i_0}=0$.
\end{itemize}
Finally; adding up (\ref{A.16}) and (\ref{A.17}) gives (\ref{024}).

\end{document}